\documentclass[12pt]{iopart}

\newcommand{\comments}[1]{}
\usepackage{iopams,graphicx} 
\usepackage{color}

\begin{document}

\title{Dynamics of topological defects in ion Coulomb crystals}

\author{H L Partner$^1$, T Burgermeister$^1$, K Pyka$^1$, J Keller$^1$ and T E Mehlst\"aubler$^1$}

\address{$^1$ Physikalisch-Technische Bundesanstalt, Bundesallee 100, 38116 Braunschweig, Germany}

\author{R Nigmatullin$^{2,3}$, A Retzker$^4$ and M B Plenio$^{2,3}$}

\address{  $^2$ Institute for Theoretical Physics and Center for Integrated Quantum Science and Technology (IQST), Albert-Einstein-Allee 11, Universit\"at Ulm, 89069 Ulm, Germany \\
		$^3$ Department of Physics, Imperial College London, Prince Consort Road, London, SW7 2AZ, United Kingdom \\
		$^4$ Racah Institute of Physics, The Hebrew University of Jerusalem, Jerusalem 91904, Givat Ram, Israel \\
		}

\ead{Tanja.Mehlstaeubler@ptb.de}

\begin{abstract}
We study experimentally and theoretically the properties of structural defects (kink solitons) 
in two-dimensional ion Coulomb crystals. 
We show how different types of kink solitons with different physical properties  can be realized, and
transformed from 
one type into another by varying the aspect ratio of the trap confinement.
Further, we discuss how impurities in ion Coulomb crystals, such as mass defects, can modify the dynamics 
of kink creation and their stability.
For both pure and impure crystals, the experimentally observed kink dynamics are 
analyzed in detail and explained theoretically by numerical simulations and calculations of the Peierls-Nabarro potential. 
Finally, we show that static electric
fields provide a handle to vary the influence of mass defects on kinks in a controlled
way and allow for deterministic manipulation and creation of kinks.
\end{abstract}

\maketitle

\section{Introduction}

Recent experimental realizations of structural defects (kinks) in ion Coulomb crystals \cite{Mielenz2013, Pyka2013a, Ulm2013, Ejtemaee2013} have raised an increasing interest in nonlinear physics in laser cooled ions, and have been proposed as a tool for storing quantum information \cite{Landa2010}.
    
Topological defects are created when a symmetry breaking phase transition is crossed non-adiabatically, such as the second-order phase transition from the linear to zigzag configuration in ion Coulomb crystals \cite{Fishman2008,Retzker2008}.
Kinks were first observed during crystallization due to laser cooling \cite{Schneider2012}.  Since then, several groups have succeeded in stable trapping of defects, either created through crystallization \cite{Mielenz2013}, or by axial \cite{Ulm2013} or radial \cite{Pyka2013a, Ejtemaee2013} quenches of the confining potential.  The storing of quasi-3D kinks and study of the interaction of different classes of defects involving mass defects are also under investigation \cite{Mielenz2013, Nigmatullin2011, LandaPC} 

We have previously used the creation of topological defects during rapid radial quenches to study the dynamics of phase transitions and to test the predictions of Kibble-Zurek theory \cite{ Pyka2013a, Ulm2013, Kibble1980, Zurek1996, delCampo2013}.
In this work, we investigate the stability and dynamics of kinks, including the influence of mass defects present in the ion crystal and the controlled manipulation of kinks using electric fields.  The dynamics occur on microsecond timescales and are difficult to resolve experimentally.  We interpret our observations at experimentally accessible timescales using simulations that provide information about the rapid dynamics and using numerical calculations of the Peierls-Nabarro (PN) potential \cite{Braun2004}, which describes the energy necessary for the kink to move within the crystal.  

The paper is organized as follows.  Section \ref{sec:background}  introduces our experimental system and describes the kink structures. Section \ref{sec:PN potential} treats the kinks as discrete solitons and introduces our method
for calculating the PN potential. Section \ref{sec:dynamics} discusses motion of kinks in the PN potential. Section \ref{sec:massdefects} investigates kinks in the presence of mass defects; in our system, heavy ions can be created by the chemical
reactions with background gas. In section \ref{sec:efield} we discuss ways in which we can create and manipulate kinks using an applied dc electric field.

\section{Creation of localized and extended topological defects} \label{sec:background}

In the experiment we use an ion Coulomb crystal of around 30 $^{172}$Yb$^+$ ions confined in a segmented linear Paul trap which is designed for trapping large crystals with minimized micromotion \cite{Pyka2013b, Pyka2013a}. Linear chains of ions
are trapped with axial frequency $\nu_{z}=24.6\pm0.5$~kHz
and radial frequencies $\nu_{x}$, $\nu_{y}\approx500$~kHz.  When the radial confinement is weakened below a critical value such that $\nu_{x,y}/\nu_z < 0.73 \times N^{0.86}$ where $N$ is the number of ions \cite{Steane1997}, the linear chain buckles into a zigzag configuration. The kinks are created by rapidly quenching the radial confinement across this transition, which causes defects to occur probabilistically \cite{Pyka2013a}. A radial asymmetry in the potential of $\nu_{y}/\nu_{x}\approx1.3$
is introduced by the end of the quench in order to confine the zigzag phase to a well-defined plane.  Reducing this ratio leads to quasi-3d kink structures \cite{Mielenz2013},
however in this work we focus on two-dimensional configurations. 

\begin{figure}
   \centering
  \includegraphics[width=\linewidth]{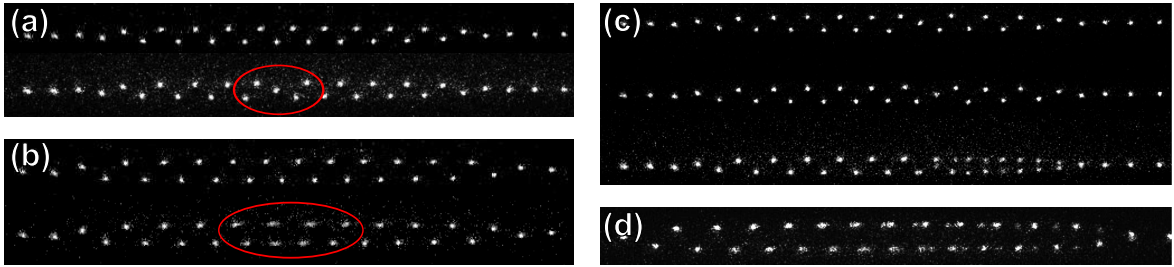} 
   \caption{ Overview of experimentally realized kinks and their dynamics in Coulomb crystals.  The kinks (indicated by red ellipses) are created during a radial quench of a crystal of about 30 ions, and imaged with exposure times of 40~ms.  (a) Zigzag and localized kink for $\nu_x / \nu_z \approx 8$.  (b)  Zigzag and extended kink for $\nu_x / \nu_z \approx 5.5$.  (c) Measured time sequence showing an odd kink leaving the chain, first moving one site and then moving out of the chain during the exposure.  The kink propagates through transverse motion of the ions.  (d) An extended kink lost from the chain during the exposure. The kink moves out of the chain through axial motion of the chains of ions on either side of the zigzag. }
   \label{fig:odd_ext}
\end{figure}

A kink can take on different configurations depending on the ratio of the transverse
and axial confinement, $\nu_{x}/\nu_{z}$. Figure \ref{fig:odd_ext} shows typical experimental images of these kink structures. For high values of $\nu_{x}/\nu_{z}$
(i.e. for small transverse inter-ion spacing) the kink has a localized structure shown
in figure \ref{fig:odd_ext}(a) which involves only $\sim3$ to $5$ ions. We
refer to this type of kink as ``odd'', using the terminology of 
\cite{Landa2010}. For low values of $\nu_{x}/\nu_{z}$ (large transverse inter-ion spacing) the kink has an extended structure as shown in \ref{fig:odd_ext}(b) that involves $\sim7$ to $10$ ions, which we refer to as an ``extended'' kink. 

To create odd kinks the radial confinement $\nu_{x}$ is relaxed from
approximately 500~kHz to 204~kHz, and for extended kinks $\nu_{x}$ is reduced from approximately 500 kHz to 140 kHz. After the quench, the spatially
resolved ion configuration is imaged at a 45 degree angle in the transverse direction onto an electron multiplying CCD camera using a typical
exposure time of 10 to 50~ms. Resulting images are evaluated for the presence
and location of kinks. The quench parameter $\tau_{Q}=t_{r}/2$, 
where $t_{r}$ is the length of the radial quench, compared with the timescale
of the secular motion $\nu_{z}\approx24.6$~kHz determines the probability of kink creation (see \cite{Pyka2013a}). 

The kinks can propagate along the chain if they obtain kinetic
energy ($\gtrsim$~5~mK) from, for example, the ramp of the radial potential (this can be on the order of tens of mK),
 collisions with background gas particles,
or a decrease in the laser cooling rate. Kinks that have moved sufficiently close
to the end of the zigzag region are lost at the edges. Ion motion happens on the timescale of the secular frequency, making kink motion impossible to resolve with exposure times of tens of milliseconds.  However, as indicated
by a superimposed kink and pure zigzag image, as in figure
\ref{fig:odd_ext}(c) and (d),  the trace of a kink can be observed moving out of the chain during the camera exposure. Experimentally, we find that the probability of observing an extended kink is higher than the probability of observing an odd kink, which suggests that odd kinks may exhibit higher losses.

In order to investigate the rapid dynamics of these losses we have performed molecular dynamics simulations, from which we observe that defects are first created as odd kinks in the early part of the quench, and transform into extended structures as the ratio of radial to axial confinement ($\nu_x/\nu_z$) is lowered further.  We observe that odd kinks are easily lost from the edge of the chain when they have enough energy to travel from site to site; in contrast, it is possible for extended kinks that are moving along the chain to remain in the crystal by oscillating back and forth and eventually stabilizing in the centre.  Since the simulations show that when driving phase transitions via radial quenches, extended kinks are always created from odd kinks, the extended kinks provide a convenient stable probe of the statistics of defect creation (in finite systems with small kink probabilities  $d <1$).  We have used this characteristic of the extended kinks to facilitate the testing of Kibble-Zurek scaling in this regime in past work \cite{Pyka2013a}.

\section{The Peierls-Nabarro potential} \label{sec:PN potential}

The PN potential is the effective potential of the kink and is defined as the energy of the system as a function of the location of the kink centre, $X$. Before the PN potential is calculated it is necessary to understand qualitatively how a
kink moves and to formulate a definition of the kink centre.  To facilitate this discussion, for a zigzag we define the inter-ion spacing in the transverse direction as $a$ and in the axial direction as $b$. 

\begin{figure} 
   \centering
  \includegraphics[scale=0.5]{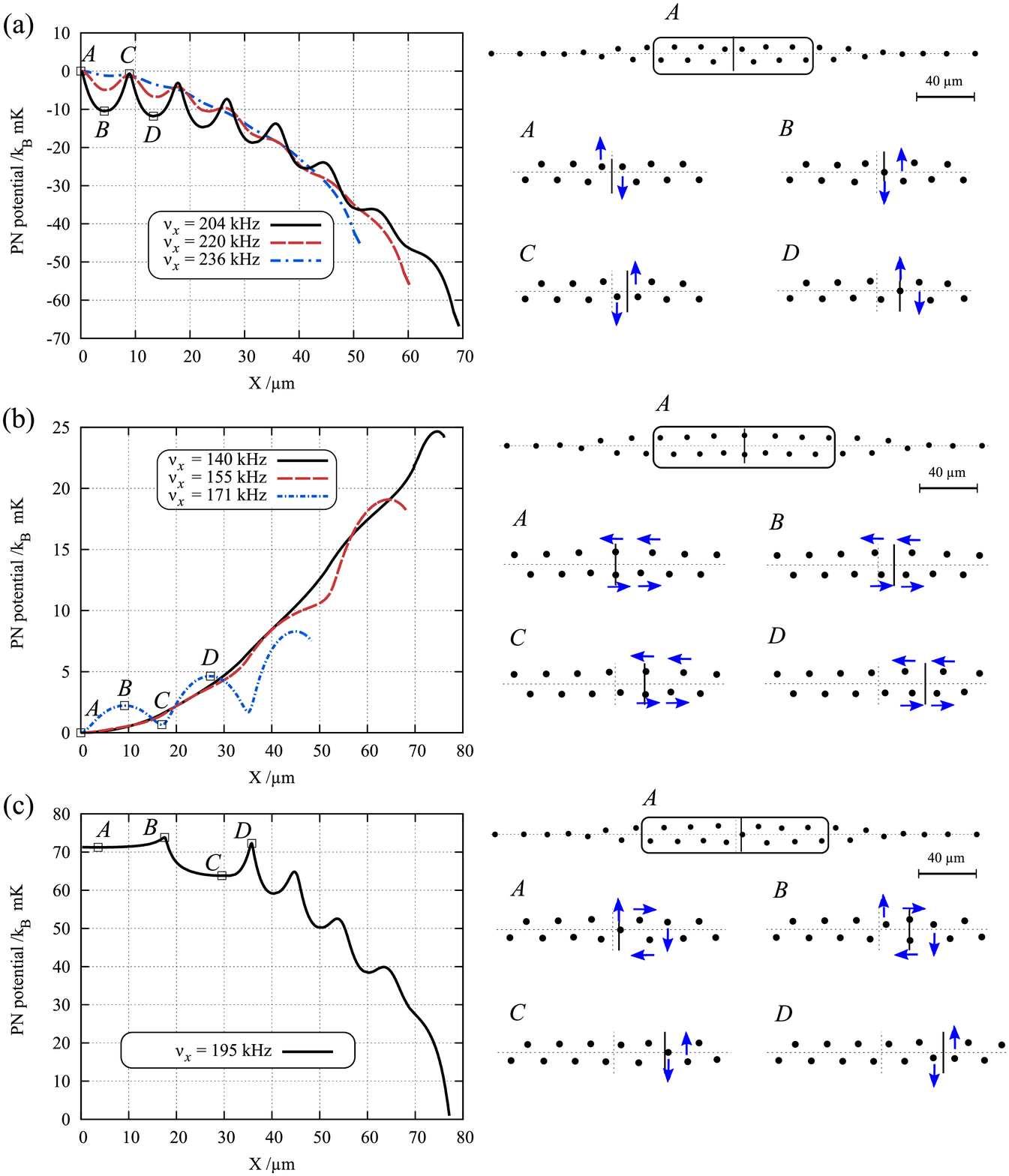} 
   \caption{The PN potentials for odd, extended, and intermediate
kink. The calculations were done for a chain of 30 Yb$^{+}$ ions
in a harmonic trap. The axial confinement was $\nu_{z}=24.6$~kHz and transverse confinement was set to a) $\nu_{x}=204$,
220 and 236~kHz for odd kinks, b) $\nu_{x}=140$, 155 and 171~kHz for extended kinks and c) $\nu_{x}= 195$~kHz for intermediate
kinks. The configurations on the right are shown with a true aspect ratio and the centre
of the kink is indicated by a solid vertical line.  The positions of the open squares on the left correspond to the vertical lines indicating the kink centre on the right, and the dashed lines indicate the centre of the crystal. }
	\label{fig:PN potentials}
\end{figure}

\emph{Odd kink. }Figure \ref{fig:PN potentials}(a) (right) illustrates
the simulated propagation of an odd kink. For this type of kink, $a<b$ and the kink propagates via the transverse
motion of the two ions that are at the interface between the two zigzag
domains. This motion is observed experimentally in figure \ref{fig:odd_ext}(c). Let $(z_{j},x_{j})$ and $(z_{j+1},x_{j+1})$ be the coordinates
of the ions at the interface between two domains. We define
the centre of an odd kink as the location where a line connecting points
($z_{j}$,$(-1)^{j}x_{j}$) and ($z_{j+1}$,$(-1)^{j+1}x_{j+1}$)
intersects with the $z$-axis. This point is marked on figure \ref{fig:PN potentials}(a) by a solid vertical line. 

\emph{Extended kink. }An extended kink has a different structure
and mode of propagation. For a zigzag chain with an extended kink $a>b$, the kink contains
an excess of one ion in either the bottom or the top row of the zigzag.
The kink is identified as the region of the excess charge density and it
propagates via the axial motion of ions, as illustrated in figure
\ref{fig:PN potentials}(b) (right). This motion is visible in the experiment in the exposure of figure \ref{fig:odd_ext}(d).  We take the kink centre as a
geometric centre of the shift in axial coordinates that a kink introduces
into the chain (see appendix A for a precise definition). 

\emph{Intermediate kink. }  As the radial confinement is decreased during a quench, the transverse separation $a$ of the
zigzag increases, to create first odd ($a<b$) and then extended ($a>b$) kink structures. In the same manner, a quench of increasing radial confinement transforms an extended kink back into an odd kink. When $a \approx b$ the propagation of the kink involves both
transverse and axial motions. Figure \ref{fig:PN potentials}(c) illustrates
this mode of propagation for a kink that occurs during the transformation between the odd
and the extended type and can appear as either an odd or extended kink at different moments during the movement. This kink structure was also experimentally observed in \cite{Ejtemaee2013}. 
The kink centre for this intermediate kink is determined using the same method as
for the extended kink. 

In order to find the PN potential for a kink in the zigzag ion crystal, we calculate the adiabatic trajectory connecting a configuration
with a kink at one end of the crystal to a configuration with the
kink at the other end. To find a minimum energy crystal configuration
with a kink at position $X$ we perform energy minimization constrained
on the kink being at position $X$ using the method of Lagrange multipliers
(see appendix A). Effectively, we start with a kink configuration
at a stationary point (a local or global minimum of the PN potential) and then
move the kink against the energy gradient using constrained energy
minimization. In this way the whole trajectory can be traced out in order to determine the PN potential. 

Calculated PN potential curves are shown in figure \ref{fig:PN potentials}
for the system of 30 $^{172}$Yb$^{+}$ ions with an axial confinement of 24.6~kHz
and (weak) transverse confinement ($\nu_{x}$) at selected values ranging from 140 to
236~kHz. First we consider the PN potential curves for the odd kinks shown
in figure \ref{fig:PN potentials}(a). It is expected that the PN potential
should be periodic in lattice spacing, since this periodicity is a
fundamental feature of discrete soliton systems. For an infinite and homogeneous system, the periodicity arises from
the invariance of a kink under translations
by an integer number of lattice spacings.  However, for a finite sized system, this
invariance is only approximately true and the form of the PN potential is modified
by the boundaries of the system. This can be seen in figure
\ref{fig:PN potentials}(a): the PN potential is periodic in lattice spacing,
but tends to decrease as the kink moves away from the centre of the chain. The height of the PN potential barriers, which indicate the minimum energy
that is required to move a kink from one lattice site to the next, increases as $\nu_{x}$ is decreased. In order for a kink to be
trapped, its kinetic energy must be lower than the height of the barrier.
During the early stages of the quench, the created odd kinks can easily
be lost, since the PN barrier is smaller than the kinetic energy in
the system and the PN gradient pushes the kink outward. As the quench
progresses, the PN barrier increases until it is eventually able to trap kinks,
which are then experimentally observable.

The PN potential curves for extended kinks are shown in figure \ref{fig:PN potentials}(b). For lower values of $\nu_{x}$, this potential is also periodic, but with a period about two
times larger than that for odd kinks, reflecting the more extended
nature of the kink structure. The PN barriers decrease with decreasing $\nu_{x}$,
since as the distance between the two rows of ions increases, the discreteness of the lattice becomes less prominent: the kink experiences
an averaged out and thus more continuous charge distribution. In this case, the effect
of the system boundaries is that the PN potential curves upward at the edges of
the zigzag region, where the charge density is higher. Thus, an extended kink tends to be pushed away from the
boundaries and trapped at the centre. When an extended kink is created from
an odd kink, it will oscillate back and forth in this PN potential
and remain trapped in the centre of the chain.  This stabilization mechanism also leads to the annihilation of pairs of extended kinks in the centre of the crystal, so that the number of observable extended kinks saturates at one defect in an 
 inhomogeneous (pure) crystal of sufficiently small size, as also noted in \cite{Mielenz2013}. 

As the odd kink transforms into an extended kink, the PN potential must transform from the periodic potential that decreases at the edges into one with a global minimum.  This happens temporally when $a \approx b$ and the defect becomes an intermediate kink, where the propagation of the kink involves the motion of ions in both the transverse and axial directions.  With this mechanism of movement, the kink propagation is such that the kink moves in steps of two lattice sites, instead of one as in the odd kink case.  This leads to the change in the periodicity of the PN potential in the intermediate stage.  In figure \ref{fig:PN potentials}(c) we see that, although this transition from odd to intermediate to extended kinks happens temporally during the quench, there is also a spatial dependence of the transformation:  at the crystal centre, intermediate and then extended kinks can form, while the kink continues to behave as an odd kink near the edges. 

These numerically calculated potentials provide an explanation for
the observed higher stability of extended kinks compared to odd kinks,
as well as the oscillating behaviour seen in the simulations. The global
minimum of the extended kink PN potential makes it more difficult for an extended
kink to escape the crystal, while in the odd case, a kink that is already moving
within the chain is easily lost as it approaches the edge of the chain. Therefore, for the
same amount of kinetic energy in the system, the average probability
for an odd kink to be lost is higher than that for an extended kink.

\section{Kink motion}  \label{sec:dynamics}

The PN potential predicts some qualitative features of the kink dynamics,
such as trapping and losses of kinks at the boundaries. By itself
the PN potential does not provide quantitative dynamical information
such as the frequency with which a kink oscillates around a local
minimum of the PN potential. To obtain such dynamic information it is necessary
to derive the effective equation of motion for the kink centre. This
can be done, for example, using the collective variable formalism (see Appendix B) \cite{Rosenau2003}. The equations of motion for kinks
typically contain an inertial term, a non-linear friction term that
arises from the coupling between kink motion and the phonon bath, and the PN force term, which
is obtained by differentiating the PN potential. Knowing the equations
of motion for kinks is important for quantitative analysis of interactions
between kink modes and phonon modes of the chain, as well as the analytic
predictions of quantities such as the frequencies of kink oscillations.
Here, we take a simple approach and use numerical simulations to measure
the oscillation frequency of an extended kink. We also use numerical
simulations to look at the response of the kink to thermal driving
of an ion chain. 

\begin{figure}
\centering
\includegraphics[scale=1.2]{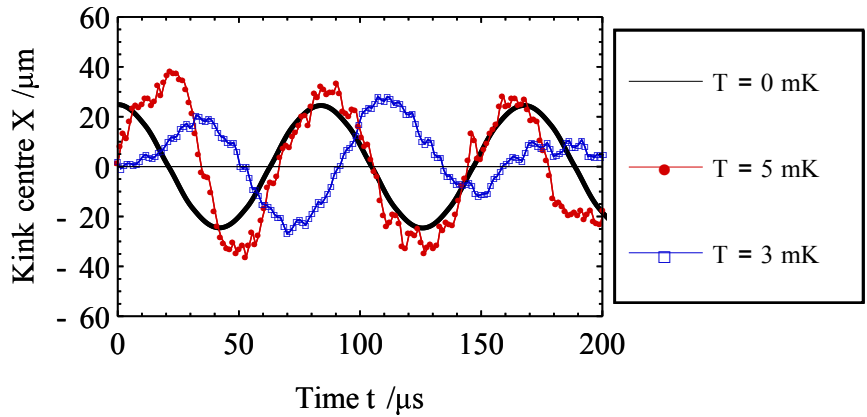}
\caption{Three examples of trajectories of kink centres obtained using numerical
simulations. The kink is of extended type in a trap with experimental parameters
$\nu_{x}=140$~kHz, $\nu_{z}=24.6$~kHz and 30 Yb$^{+}$
ions. The temperatures were set to 0~mK (black line), 5~mK (red line)
and 3~mK (blue line). The finite temperature results were obtained
from Langevin dynamics simulations with a friction coefficient of $\eta=3\times10^{-21}$ due to laser cooling forces (see \cite{Metcalf1999}).  For an example of an experimental image of an oscillating kink, see figure \ref{fig:drag}(c)(iii).  \label{fig:dynamics}}
\end{figure}

We consider an extended kink of 30 Yb$^{+}$ ions in a trap with $\nu_{x}=140$~kHz and $\nu_{z}=24.6$~kHz. The PN potential for this kink is shown
in figure \ref{fig:PN potentials}(b). Molecular dynamics simulations show that these kinks
oscillate about $X=0$, which is the point of global minimum of the
PN potential. To observe this oscillation in its pure form, we initialize
the positions of ions in a configuration taken from the adiabatic
trajectory, set the initial ion velocities to zero, and numerically
evolve the equations of motion without damping or stochastic forces.
Once ion trajectories have been found, the position of the kink centre
at each time instance is calculated using equation \ref{eq:app3}.
For initial conditions $X(0)=25$~$\mu$m and $\dot{X}(0)=0$ the
trajectory of a kink is shown in figure \ref{fig:dynamics}(a) ($T=0$~mK line).  The oscillation is very regular with no noticeable damping, which indicates weak coupling of the kink to the phonon modes of the crystal.
The frequency of the oscillation is 11.9~kHz. 

Next, we consider the excitation of the motional kink mode by
thermal noise. The thermal dynamics of the chain is modelled using
the Langevin equation, where the effect of the thermal bath is modelled
by the addition of frictional and stochastic forces whose statistical
properties satisfy the fluctuation dissipation theorem (see \cite{Pyka2013a}
for more details concerning the simulation approach). The kink is
initialized at $X=0$ and its evolution is calculated numerically
for the two temperatures $T=3$~mK and $T=5$~mK. A 400~$\mu$s sample
interval of the thermalized evolution of kink centre is shown in
figure \ref{fig:dynamics}. The kink oscillates with a frequency of
$\approx12$~kHz, and the amplitude is larger for the crystal at higher
temperature as expected. These simulations allow us to estimate the timescales and amount of heating necessary to move kinks within our ion Coulomb crystal.  

\section{Mass Defects}\label{sec:massdefects} 

Next we consider how an impurity in the Coulomb crystal influences the kink solitons.  A mass defect in the ponderomotive potential of a Paul trap experiences a different radial confinement than the rest of the ions, since the secular frequency $\nu_x \sim 1/m$ where $m$ is the mass \cite{Ghosh1995}.  
Because of this, the mass defect changes the configuration of the crystal and deforms the PN potential, consequently affecting the dynamics of the kinks. This modification of the PN potential depends on the location and the mass of the impurity species and the type of kink.  
We primarily consider extended kinks with mass defects heavier than the surrounding $^{172}$Yb$^+$  ions in this analysis.  

\begin{figure}[h]
   \centering
  \includegraphics[width=\linewidth]{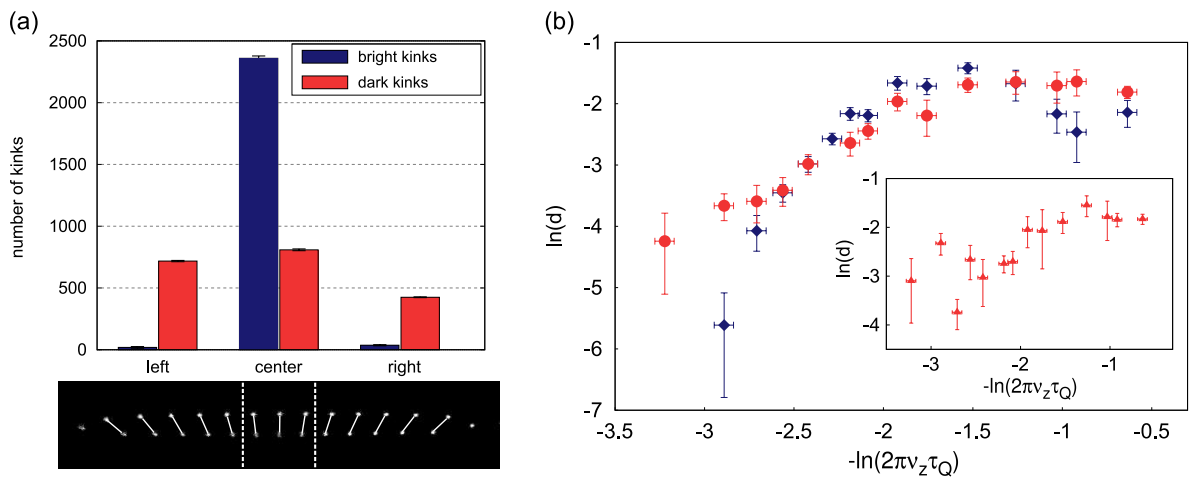} 
   \caption{Influence of molecules on spatial distribution and creation of kinks.  (a) \emph{top}: Histogram of the experimentally observed kink location with and without molecules.  Off-centre kinks were rarely observed in pure crystals.  With molecules present, the probability to have an off-centre kink increases.  This indicates trapping of kinks by molecules. 
   \emph{bottom}: Definition of the location of a kink as left, right, or centre.  The centre of the kink must be outside of the middle three sites to be considered off-centre. (b) Scaling of measured kink density with the quench parameter $\tau_Q$, with and without one molecule present in the chain.  Losses at slow ramps are reduced when a molecule is present.  \emph{inset}: Scaling of kink density for two molecules in the chain.}
   \label{fig:mol_vs_bright}
\end{figure}

In our experiment, heavy mass defects are implanted by the creation of molecules during collisions of the $^{172}$Yb$^+$ ions in the crystal with background gases. In a fluorescence image of the chain, a molecular mass defect appears as a dark ion.  The $^{172}$Yb$^+$ ions can react with water to produce YbO$^+$, Yb(OH)$^+$, and other molecules \cite{Sugiyama1995, Sugiyama1997, Rutkowski2011}. By performing parametric excitation of two to four ions including a mass defect, we have measured primarily YbO$^+$ or Yb(OH)$^+$; the addition of hydrogen was not resolvable by this method.  These heavy masses distort the crystal as well as the kink configurations.  The amount of distortion can be used as an alternative method to estimate the mass of the molecule by comparing the experimental images with crystal configurations inferred from molecular dynamics simulations.  This analysis indicates the presence of even heavier molecules with one to three oxygen atoms involved.   In the simulations, we investigate mass defects with representative masses $m =$~188 (YbO$^+$) and $m =$~220 (YbO$_3^+$) (all masses are given in atomic mass units), consistent with observations in our system.  Figure \ref{fig:drag}(a) shows three examples of structural defects including a molecule.  The first image is of a crystal containing a light molecule (e.g. YbO$^+$), while the greater distortion evident in the other two images indicates the presence of a heavy molecule (e.g. YbO$_3^+$).  In the experiment, the position of mass defects within the crystal is not controlled and varies over the crystal.

Experimentally, we see that the spatial distribution of kink positions within the crystal is altered in the presence of a mass defect.  This indicates that a mass defect can stabilize a kink at a different location than the centre of the crystal. Figure \ref{fig:mol_vs_bright}(a) shows a histogram of kink location with and without molecules present for, in each case, about 2000 observed extended kinks. 
In the case with molecules, the appearance of defects in the left and right sides of the crystal becomes much more probable.  This indicates that kinks which would otherwise either escape the chain or settle in the centre of the PN potential after oscillating are instead stopped at the molecule. 

We also see evidence that the probability of kinks to be created  (i.e. the kink density $d$) and trapped during the radial quench is modified by the presence of molecules. Figure \ref{fig:mol_vs_bright}(b) shows a comparison of measured kink density as a function of the quench parameter $\tau_Q$ for crystals with mass defects compared with the case of pure crystals. 

\subsection{Peierls-Nabarro potential with impurities}
\begin{figure}
   \centering
  \includegraphics[scale=0.5]{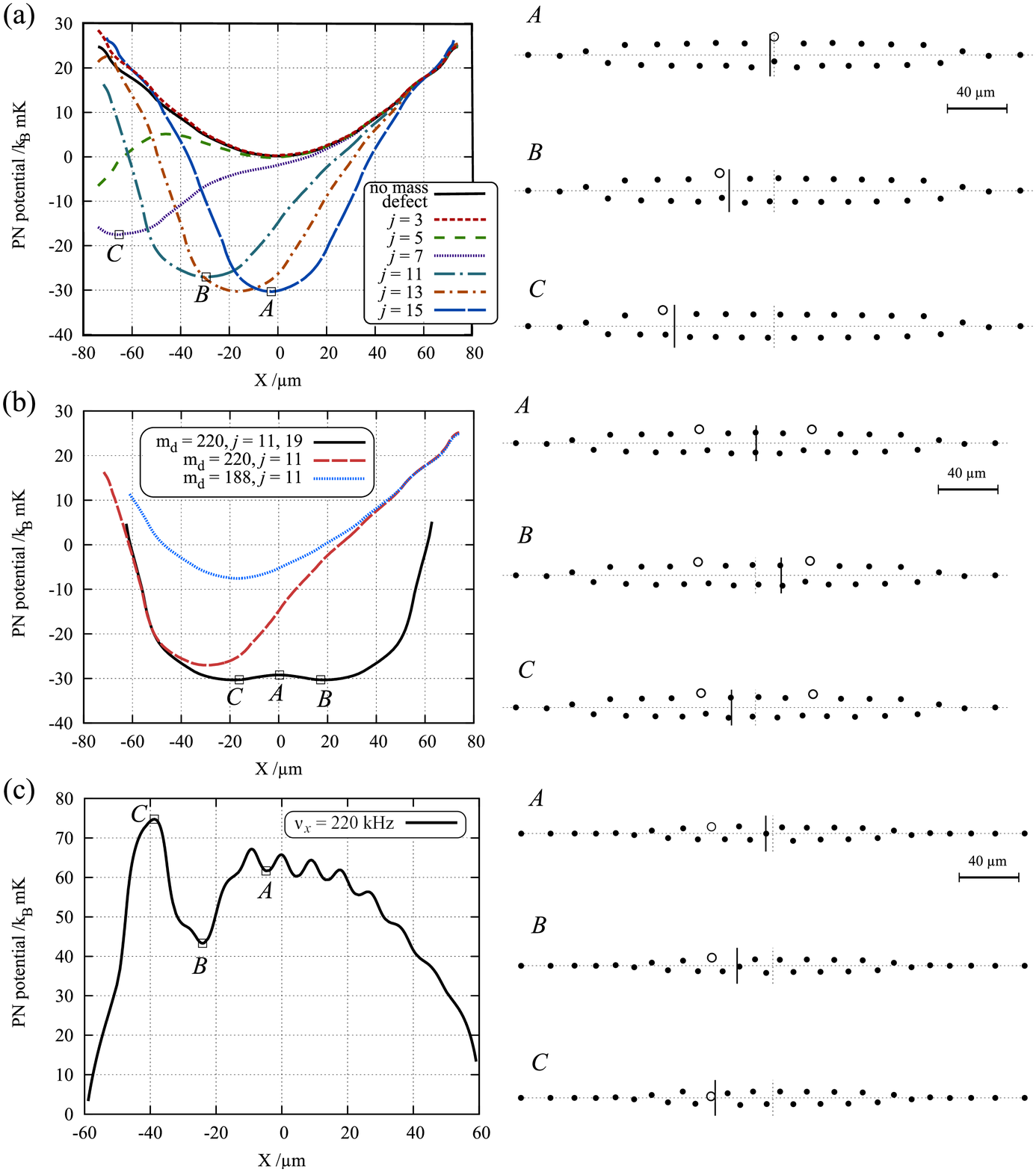} 
   \caption{Kink-confining PN potentials for crystals with mass defects. (a) PN potential calculated for the extended kink case with a  molecule of mass $m = 220$  located at different positions in the crystal  ($\nu_z=$~24.6~kHz and $\nu_x=$~140 kHz). The losses observed in the simulation at position 5 are explained by the low barrier to leave the chain. (b) Comparison of PN potentials with mass defects of $m =$~220 and $m =$~188. The depth of the PN potential increases with the mass of the defect. Also shown is the case of two heavy molecules located at positions 11 and 19 (black line) creating a double well potential.  (c) PN potential for an odd kink with transverse trapping frequency $\nu_x=$ 220~kHz ($\nu_z=$~24.6 kHz) and a molecule with $m=220$ at position 11.  The positions of the open squares on the left correspond to the vertical lines indicating the kink centre in the configurations on the right. }
   \label{fig:PN potential_molecule}
\end{figure}

The modified spatial kink distribution can be understood by evaluating the PN potential in the presence of mass defects.  Figure \ref{fig:PN potential_molecule} shows the PN potential for kinks in a crystal with a molecule at fixed positions.  These PN potentials have a local minimum at the mass defect.  Figure \ref{fig:PN potential_molecule}(a) and (b) show that for extended kinks, the local minimum approaches a global minimum as $\nu_x$ is lowered, whose axial location in the chain is modified from the centre to the location of the molecule.  This is the case we have investigated experimentally. Since the molecule is less bound in the radial potential than the rest of the ions, it is easier for the molecule to escape into the transverse plane, thus lowering the charge density and the potential energy.  This allows stable trapping of kinks at the location of the mass defect.  Moreover, in the case of more than one mass defect, the creation of local minima at the positions of the mass defects allows trapping of more than one extended kink in such mixed species crystals, despite their small size (see black line in figure \ref{fig:PN potential_molecule}(b).  Experimentally, we observed the stable trapping of two extended kinks in the presence of two heavy molecules (see figure \ref{fig:drag}(b)).
Finally, we also calculate the PN potential for ``odd" kinks in the presence of molecules.
In this case, shown in figure \ref{fig:PN potential_molecule}(c), the molecule also induces a deep local minimum that can trap the kink before it leaves the chain.  For both odd and extended kinks, the mass defect can therefore stabilize a kink nearby to its position.

\subsection{Kink creation with impurities}
\begin{figure} 
   \centering
  \includegraphics[width=\linewidth]{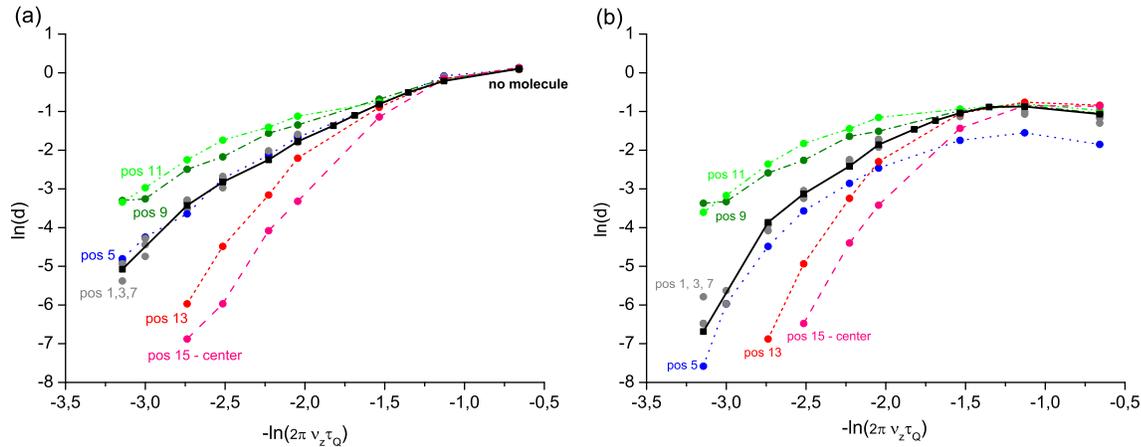} 
   \caption{Simulated kink density scaling with ramp time for a molecule of mass $m =$~220 at various positions in a chain of 30 ions. (a) Density of kinks created during the ramp.   (b) Density of kinks still present 400~$\mathrm{\mu s}$ after the ramp started and energy is dissipated (experimentally accessible kinks). }
   \label{fig:molecule_positions}
\end{figure}

In order to investigate the influence of mass defects on kink creation, the dynamics of the phase transition must be considered.  This includes an interplay of the propagating phase front velocity and the speed of sound \cite{Pyka2013a, delCampo2010, deChiara2010, Zurek2009, Sabbatini2012}.  In addition, the weaker radial confinement of the mass defect causes the phase transition to set in at an earlier time at the position of the molecule.  These effects introduce a dependency of the defect probability on the quench rate. The inhomogeneous nature of the crystal due to the harmonic trap confinement also leads to an expected dependence on the position of the mass defect. For this reason it becomes necessary to consider kink creation as a function of both the quench rate and the mass defect position.  In our numerical simulations we typically vary the quench parameter $\tau_Q$ from 15 to 100~$\mu$s and deduce the kink density with mass defects placed at fixed positions.    A simulation of the density of extended kinks with a molecule of mass 220 at various positions along the ion chain for different ramp times is shown in figure \ref{fig:molecule_positions} for a crystal of 30 ions.  The position numbers indicate the number of ions counted from the left, where the leftmost ion is at position zero.  Because of symmetry only half of the positions need to be simulated.  Two cases are shown: first, including all kinks that are created during the ramp (figure \ref{fig:molecule_positions}(a)), and second, kinks that are still present, and therefore measurable, 400~$\mathrm{\mu s}$ after the ramp when a stable kink density is reached (figure \ref{fig:molecule_positions}(b)).  For each case, the kink density without molecules is plotted for comparison (black line).

First, we consider the case of the created kinks in figure \ref{fig:molecule_positions}(a).  For fast quenches the molecule has no effect on kink creation, but for slow quenches, the kink density exhibits a strong position dependence.  When the molecule is in the centre (e.g. positions 13 or 15), kink formation is suppressed for slow radial quenches. Because the molecule is less bound radially than the rest of the crystal,  it  breaks out of the linear chain earlier and initiates the phase transition sooner than in a pure crystal. This leads to a prolonged time period in which the phase transition spreads out, suppressing defect formation for sufficiently slow quenches \cite{delCampo2010, deChiara2010}. At intermediate sites (e.g. positions 9 and 11), the kink density is enhanced. Here, in addition to the quench induced phase transition, which always begins at the crystal centre due to its inhomogeneity, a second phase transition is initiated at the molecule's position.  This occurs because of the lower radial confinement of the mass defect and the higher charge density near to the centre pushing the molecule out of the chain. These two phase transitions take place independently,  bringing about an enhanced probability of kink formation due to the possible conflict when the phase fronts meet. With the molecule at the edges of the chain (positions 1-7), the kink density is unaffected by the presence of the molecule. In the outer parts of the crystal the molecule does not experience a sufficient repulsion in this region of lower charge density to initiate the independent phase transition. The molecule position's influence on kink creation behaviour is reduced at faster ramp times and eventually has no effect (i.e. for $\tau_Q<$~30~$\mu$s). In this regime the system behaves as a homogeneous system (the phase front is faster than the speed of sound), and this characteristic dominates the kink creation dynamics.

The case of experimentally accessible kinks is shown in figure \ref{fig:molecule_positions}(b). The density of kinks remaining after 400~$\mu$s is mainly governed by the altered rate of kink creation shown in figure \ref{fig:molecule_positions}(a), with two exceptions.  First, in positions 9 to 11, the losses from the crystal are strongly reduced for slower ramps.  For these intermediate positions, 
the PN potential at the molecule provides a very strong confinement that traps kinks formed at the centre as they start to move out of the crystal, reducing losses. Second, at position 5, high losses occur.  When the molecule is around position 5, it is easy for the kink to be lost due to the reduced barrier of the PN potential caused by the molecule, as seen in figure 5(a).


\section{Mass defects with electric fields} \label{sec:efield}

Finally, we discuss the influence of an additionally applied electric dc field on the creation and dynamics of topological defects in a crystal containing mass defects. A heavier ion experiencing a weaker radial confinement than the $^{172}$Yb$^+$ ions is displaced farther from the trap axis by an external field. The radial displacement relative to the lighter ions can therefore be controlled, causing the mass defect to behave as a molecule with increased effective mass.   The continuous tunability of the effective mass of the molecule provides a rich playground of methods to influence kink creation and behaviour. In this section we discuss several techniques for extended kink manipulation via the application of electric fields.

\begin{figure}  
   \centering
  \includegraphics[scale=0.9]{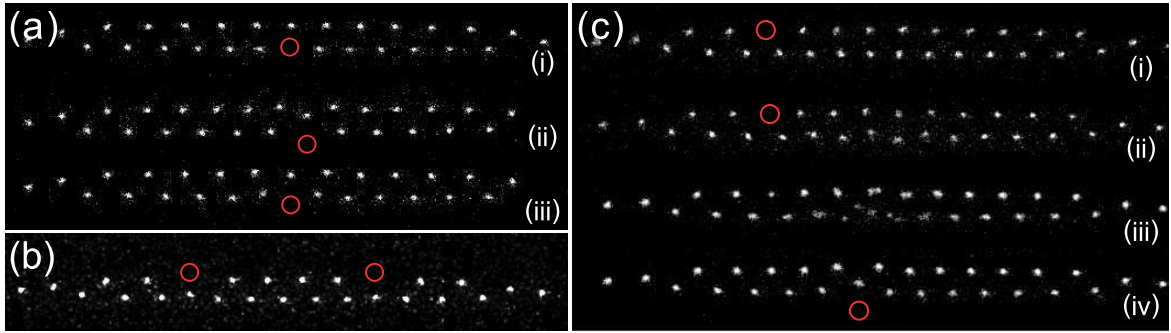}
   \caption{Experimental images of kinks and kink dynamics with molecules (indicated by the red circles).   (a)(i) Extended kink with a light  molecule (m~$\approx$~188).  (ii) Extended kink with a heavy  molecule (m~$\approx$~220) located at the kink. (iii) Extended kink with the heavy molecule moved by one lattice site. Comparison with numerical simulations shows that the distortion in the crystal depends on the mass and lattice site of the molecule.  
(b) Two extended kinks trapped at two molecules in a single crystal. (c) Time series demonstrating control of a kink and molecule using an electric field (20~ms exposures).  Image (i) shows a kink in the center of the crystal after the radial quench.  Image (ii) shows the crystal after the adiabatic application of an electric field in the plane of the zigzag (downward in the image).  In (iii) the laser cooling light is shifted a few MHz 
closer to resonance, heating the Coulomb crystal and providing the kink with motional energy.  In this image, motion during the exposure is visible when the kink begins to oscillate in the PN potential, gets trapped by the mass defect, and returns to the centre with the molecule. The kink and molecule then stabilize (iv).  } 
   \label{fig:drag}
\end{figure} 

First we consider the case of applying a time-independent electric field in the $x$-direction during a quench of the radial potential. Here, in the linear phase the molecule is already displaced from the linear chain of ions in the preferred direction of the electric field. The amount of displacement depends on the mass of the molecule and the amplitude of the electric field. Since the molecule is already displaced from the linear chain, it initiates the phase transition as soon as the quench is applied.  This leads to an amplification of the position-dependent suppression and enhancement of kink formation discussed in the previous section for the molecule without an external field.  As the radial confinement is lowered, not only is the mass defect displaced further from the crystal, but the entire ion chain moves to a new equilibrium position. Compared to the case without additional fields these rapid movements increase the kinetic energy in the system by approximately one order of magnitude, depending on the quench time $\tau_Q$ and the strength of the electric field.  

Next we describe a method for using the PN potential to control the dynamics by moving a molecule together with a kink to the centre of the crystal. To do this, we first apply the radial quench without an additional field to produce a kink and subsequently apply the dc electric field. This method circumvents the problem of excitations due to a fast translation of the whole crystal. In simulations we have observed that after the ramp, the PN potential minimum at the position of the molecule causes a passing kink in the presence of laser cooling to be trapped at the molecule. The emerging kink configuration pushes the molecule farther outside of the ion crystal (as in figure \ref{fig:drag}(a)). When the displacement of the molecule reaches a critical point,  the influence of the overall axial trapping potential becomes stronger than the interaction with the neighbouring ions, causing the molecule to move to the centre. Because the PN potential minimum is always located at the position of the molecule (see figure \ref{fig:PN potential_molecule}(a)) the kink is dragged by the mass defect to the centre of the crystal. The field amplitude that is needed to induce this effect depends on the molecule mass and position. In simulations we found that the behaviour sets in at electric field values from $\approx$~10~V/m ($m =$~220) to $\approx$~100~V/m ($m =$~188), in contrast to the case without a kink involved, where approximately twice as much field strength is required to move the molecule to the centre of the chain. We also observed a position dependence due to the inhomogeneity of the crystal: defects near the ends of the zigzag region require a higher electric field strength than molecules near the center in order to trap a kink moving with the same kinetic energy. 

Although the dynamics occur at microsecond timescales, we can induce and observe this behaviour experimentally in steps. Figure \ref{fig:drag}(c) shows an example of the experimental sequence with a molecule of $m \approx$ 188.  After a quench of length 2$\tau_Q$, a series of pictures is taken with exposure times of 20~ms, with 20~ms between exposures. After a kink has been created by the ramp (i), we increase the electric field linearly to $\approx 110$~V/m over 60~ms (ii). In (iii) we heat the entire ion chain by reducing the detuning of the cooling laser by a few MHz. 
This causes the kink to oscillate in the confining PN potential, similarly to what is shown in figure \ref{fig:dynamics}, until it reaches the position of the molecule.  As the kink gets trapped at the molecule it pushes the molecule further out of the crystal where it is free to move to the centre of the axial potential. When the molecule changes its position, the PN potential minimum and the kink trapped therein move along with it.  The kink and molecule finally stabilize at the centre of the crystal (figure \ref{fig:drag}(c)(iv)).  We have verified that for the parameters used, the molecule does not move in the crystal without the presence of a kink.

\begin{figure}
   \centering
  \includegraphics[scale=0.4]{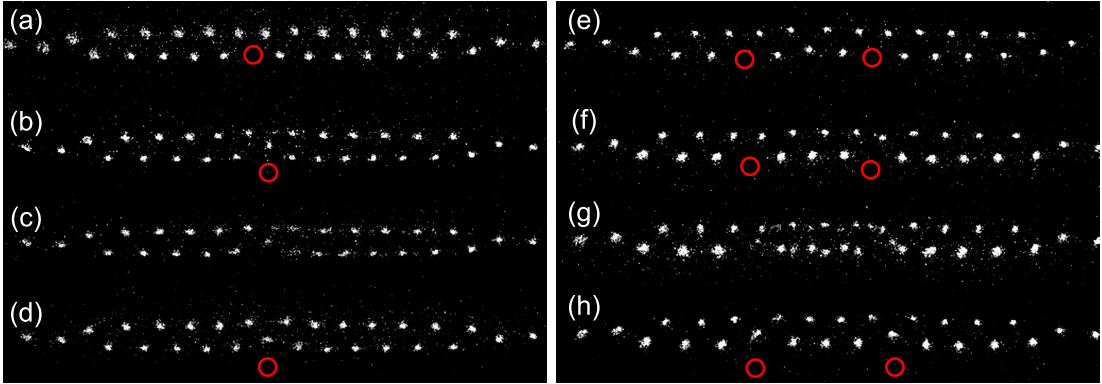} 
   \caption{(a-d): Creation of one kink defect using an applied electric field. (a) Ion crystal in the zigzag phase with one defect.  (b) When the electric field is applied (downward and in the plane of the zigzag), the molecule is displaced and two extended kinks are visible; one at the location of the mass defect, the other just to the right of it.  (c) The kink on the right is lost out of the side of the crystal. (d) Only the kink at the mass defect remains.  (e-h): Two kinks created in a crystal with two mass defects. (e) Ion crystal in the zigzag phase with two mass defects.  (f) As the electric field is increased, the chain is distorted. On the right side, an ion moves toward the vacancy left by the molecule as it is pulled downward by the electric field.  (g) A kink and anti-kink are created by the right molecule.  (h) The two kinks stabilize at the locations of the molecules. } 
   \label{fig:kinkcreation}
\end{figure} 
 
Last we look at two techniques that could be used for deterministic kink creation that we have explored using simulations, and have also been observed in the experiment. First, using a time-independent electric field during the radial quench, we have the possibility for two molecules in the appropriate positions to create kinks deterministically, because the two molecules always move in the same preferred direction when entering the zigzag phase.  For example, with an even (odd) number of ions between the molecules in the linear chain, a kink must (must not) form.  Simulations indicate that an external electric field of 10~to~20~V/m  is sufficient to produce this effect.

The second method is independent of the molecule position and starts with a crystal in the zigzag phase with a single mass defect. No quench of the trapping potential is necessary to produce the kink. To create a kink we ramp up the electrical field. Again, as the field increases the molecule is increasingly displaced from the zigzag, until it is sufficiently far outside the chain that it becomes decoupled and moves toward the centre of the crystal in the axial trapping potential. When the heavy ion leaves its position in the chain it produces two kink defects. One of them behaves like a kink in a chain without molecules and starts to move in the confining PN potential. The other one is directly located at the molecule and together they move to the centre of the chain, because the axial trapping potential dominates the motion of the molecule. The production of the two defects can be understood more easily in terms of lattice sites. As the molecule leaves its original lattice site there is a vacancy created in the lattice. This unstable vacancy then produces the first kink. The latter defect is created, because there is no available lattice site for the molecule.  It can be interpreted as an interstitial defect, which then converts into another kink. This can also be seen as the production of a kink and anti-kink.  Figure \ref{fig:kinkcreation} shows two experimental time series demonstrating this method applied to crystals with one and two mass defects.

\section{Conclusion} \label{sec:conclusion}

We have investigated the confining Peierls-Nabarro potentials of 
localized and extended kink solitons in two-dimensional ion Coulomb crystals.
Each type of kink can be transformed into the other type 
by varying the aspect ratio of the trap potentials. 
Further, we have shown that a heavy mass defect can trap and stabilize a kink at its
location by deforming the PN potential and creating local minima. This allows trapping of more than
one extended kink soliton in the Coulomb crystal and offers the possibility to study 
kink-kink interaction in the future. The presence of mass defects displays a complex dynamics during the phase transition, leading to enhanced and suppressed  kink formation.   
In our experiment we have used molecules to study the influence of mass defects; however, for a more deterministic control of mass defects, another addressable ion species can be used.

Tuning this influence of mass defects with controlled electric dc fields 
offers a rich playground for the manipulation of kink solitons.
In particular, we have shown first evidence of kink solitons being deterministically 
created by electric field ramps and the manipulation of an extended kink attached to a mass defect.  Since the extended defects can be transformed back into localized kinks, these techniques could also be applied to other various types of kink solitons. 

The techniques that we have established here help to open up the field of study in the rich physics of kink solitons in ion Coulomb crystals. Besides the dynamics and interaction of multiple kinks, this may also include the study of internal vibrational modes of kinks \cite{Mielenz2013, Landa2010} and ultimately, the quantum dynamics of kinks and quantum statistical mechanics in ion chains \cite{Retzker2008}.



\appendix

\section{Calculation of the Peierls-Nabarro potential}

In this section we explain in detail the procedure that was used to
calculate the PN potentials. First we define the function used to calculate
the kink location of the kink centre. Then the problem of finding the adiabatic trajectory and hence 
PN potential is formulated as an optimization problem that can be solved using the method 
of Lagrange multipliers.

\subsection{Definition of kink centre}

Consider a harmonic trap such that a single charged particle of mass
$m$ would oscillate in it with frequencies $\omega_{x}$ and $\omega_{z}$.
The potential energy of the system of $N$ particles trapped in such
a trap is given by

\begin{equation}
V=\sum_{j=1}^{N}\frac{1}{2}\frac{m^{2}}{m_{j}}\left(\omega_{x}^{2}x_{j}^{2}+\omega_{z}^{2}z_{j}^{2}\right)+\frac{e^{2}}{4\pi\epsilon_{0}}\sum_{i<j}^{N}\frac{1}{\sqrt{(x_{j}-x_{i})^{2}+(z_{j}-z_{i})^{2}}},\label{eq:app1}
\end{equation}
where $m_{j}$ is the mass of $j$th ion. 

Suppose the charges are distributed in a zigzag kink configuration;
the $z$-axis being the longitudinal direction and $x$-axis
the transverse direction. We denote the charge configuration by a
vector $\textbf{Q}=(z_{1},\ldots z_{N},x_{1},\ldots x_{N},)$, where
the ions are numbered such that $z_{1}\leq z_{2}\ldots\leq z_{N}$.
The centre of the kink, $X$, is a function of $\textbf{Q}$ i.e.
$X=g(\textbf{Q})$. For odd kinks we choose

\begin{equation}
g(\textbf{Q})=\frac{-(-1)^{j+1}x_{j+1}z_{j}+(-1)^{j}x_{j}z_{j+1}}{(-1)^{j}x_{j}-(-1)^{j+1}x_{j+1}},\label{eq:app2}
\end{equation}
with $x_{j}x_{j+1}>0$ i.e. the kink is located between the $j$th
and $(j+1)$th ion. For extended kinks we choose

\begin{equation}
g(\textbf{Q})=\frac{\sum_{j}^{N-1}\frac{1}{2}\left(z_{j}+z_{j+1}\right)\left[\left(z_{j+1}-z_{j}\right)-\left(z_{j+1}^{(0)}-z_{j}^{(0)}\right)\right]^{2}}{\sum_{j=1}^{N-1}\left[\left(z_{j+1}-z_{j}\right)-\left(z_{j+1}^{(0)}-z_{j}^{(0)}\right)\right]^{2}},\label{eq:app3}
\end{equation}
where $z_{j}^{(0)}$ denotes the $z$ coordinate of an ion in chain
in a configuration with no kink. Equation (A.3) is effectively a weighted average. It  measures the location along the $z$-axis of the centre of charge distribution distortion that is introduced by the presence of the kink.

\subsection{Evaluation of adiabatic trajectory}

The adiabatic kink trajectory is defined as a trajectory that takes
a kink at position $X_{1}$ to position $X_{2}$ with minimum amount
of work. Let $\textbf{f}(X)$ denote the configuration of particles
corresponding to a kink at position $X$ along the adiabatic trajectory.
By definition $\textbf{f}(X)$ has to minimize the energy function
whilst satisfying $g(\textbf{f}(X))=X$. Thus $\textbf{f}(X)$ is
found by solving the following optimization problem

\begin{eqnarray}
\textrm{minimize }V(\textbf{Q})\label{eq:app4}\\
\textrm{subject to }g(\textbf{Q}) & = & X.\nonumber 
\end{eqnarray}
This can be done using the method of Lagrange multipliers. First we
construct a Lagrange function $\Lambda(\textbf{Q},\lambda)$ 

\begin{equation}
\Lambda(\textbf{Q},\lambda)=V(\textbf{Q})+\lambda(g(\textbf{Q})-X),\label{eq:app5}
\end{equation}
where $\lambda$ is an additional variable called the Lagrange multiplier.
The solution to optimization problem (\ref{eq:app4}) is obtained
by solving the following set of simultaneous equations

\begin{equation}
\nabla\Lambda(\textbf{Q},\lambda)=0,\label{eq:app6}
\end{equation}
where $\nabla\equiv\left(\partial/\partial x_{1},\ldots,\partial/\partial x_{N},\partial/\partial z_{1},\ldots,\partial/\partial z_{N},\partial/\partial\lambda\right).$
Equation (\ref{eq:app6}) can be solved numerically by using iterative
gradient descent methods. It is best to select an initial guess that
is close to the true solution in order to not get trapped in a wrong
local minimum during the implementation of gradient descent algorithm.
Typically, we take a kink configuration in a known stable stationary
configuration $\textbf{f}(X_{0})$ and use it as an initial guess
for finding $\textbf{f}(X_{0}+\delta X)$, where $\delta X$ is small.
Once $\textbf{f}(X_{0}+\delta X)$ is found, we use it as an initial
guess for finding $\textbf{f}(X_{0}+2\delta X)$ and so on. This way
the whole adiabatic trajectory of a kink can be traced out.

The potential energy of a kink along the adiabatic trajectory is the
Peierls-Nabarro potential $U(X)$ i.e.

\begin{equation}
U(X)=V(\textbf{f}(X)).\label{eq:app7}
\end{equation}
Note that if we know that $U(X)$ monotonically decreasing between
some two points $X_{1}$ and $X_{2}$ then the adiabatic trajectory
can simply be evaluated by initializing a kink at point $X_{1}$ and
calculate its dynamics under heavy damping. The kink will ``slide
downhill'' along the adiabatic trajectory from the point $X_{1}$
to the point $X_{2}$. Where as Newton's laws always move the kink
along the energy gradient, the constrained optimization can be used
to push the kink with or against the energy gradient.

\section{Equation of motion of the kink}
Here we explain how the equation of motion of the kink can be derived using 
the collective-variable formalism. Some of the earliest work developing this 
approach was in the studies of solitons in the Klein-Gordon quantum field theories
\cite{PhysRevD.11.2943,PhysRevD.12.1678}. The theory was developed further in the context of
discrete kinks \cite{PhysRevB.38.6713} and used in a number of works analyzing the 
dynamics of discrete kinks such as Sine-Gordon kinks \cite{PhysRevB.33.1904,PhysRevB.33.1912,PhysRevB.39.361,PhysRevB.40.2284}.

The first step in the derivation of the equation of motion is to write
the vector that denotes the state of the system as a sum of two parts:
a contribution from the kink configuration taken from the adiabatic
trajectory and an additional perturbation

\[
\textbf{Q}=\textbf{f}(X)+\textbf{q}.
\]
We may refer to the configuration $\textbf{f}(X)$ as a bare kink
configuration and the vector $\textbf{q}=(q_{1},\ldots,q_{N})$ as
kink dressing, which may come for instance from the excited phonon
modes of the chain. In the collective variable formalism the variable,
$X$, is promoted to a dynamical variable and a corresponding canonical
momentum, $\Pi$, is introduced. Thus to derive the equations of motion
one must perform a canonical transformation from the original $4N$
variables $Q_{j}$, $P_{j}$ to the new $(4N+2)$ variables $q_{j}$,
$p_{j}$, $X$ and $\Pi$. Since the number of degrees of freedom
is increased by two one has to introduce two constraints. A good choice
of the constraints is 

\begin{eqnarray}
C_{1} & = & \textbf{f}(X)\cdot\textbf{q}=0,\label{eq:C1}\\
C_{2} & = & \textbf{f}'(X)\cdot\textbf{q}=0,\label{eq:C2}
\end{eqnarray}
where prime denotes differentiation with respect to the argument e.g.
$\textbf{f}'(X)=\left(\partial f_{1}(X)/\partial X,\ldots,\partial f_{2N}(X)/\partial X\right)$.
Constrained (\ref{eq:C1}) and (\ref{eq:C2}) are chosen in order
to minimize $q_{j}$ in the vicinity of the kink. The canonical transformation
was performed in \cite{PhysRevB.33.1904} and it was also proven that the result
can be obtained using a simpler projection operator approach \cite{PhysRevB.38.6713}.
Here, we quote the result

\begin{eqnarray}
M(X)\ddot{X}+\sum_{j}f'_{j}\frac{\partial V}{\partial q_{j}}+\textbf{f}\cdot\ddot{\textbf{q}}+\textbf{f}'\cdot\textbf{f}''\dot{X}^{2} & = & 0 \label{eq:motion1}\\
\sum_{n}\left(\delta_{ln}-\mathcal{P}_{ln}\right)\left(\ddot{q}_{n}+f_{n}''\dot{X}^{2}+\frac{\partial V}{\partial q_{n}}\right) & = & 0 \label{eq:motion2}
\end{eqnarray}
where $M(X)\equiv\textbf{f}(X)\cdot\textbf{f}(X)$, $\mathcal{P}_{ln}=f'_{l}f'_{n}/M$.

In deriving equations (\ref{eq:motion1}) and (\ref{eq:motion2}) no approximations are made, thus a solution
of (\ref{eq:motion1}) and (\ref{eq:motion2}) should give the kink dynamics, which is identical to
the dynamics obtained by solving Newton's laws of motion. The collective-variable equations can give a better
insight into kink dynamics since they explicitly keep track of the motion of the kink. They can serve as a starting point for predicting dynamical features such as, for example, the PN oscillation frequency and diffusion coefficients of kinks.

\ack
We thank Piet O. Schmidt for comments on the manuscript. This work was supported by the EU STREP PICC, the Alexander von Humboldt Foundation (M.B.P.), by EPSRC (R.N.) and by DFG through QUEST.

\section*{References}
\bibliographystyle{unsrt}
\bibliography{KDbib}

\begin{thebibliography}{10}

\bibitem{Mielenz2013}
Mielenz M, Brox J, Kahra S, Leschhorn G, Albert M, Schaetz T, Landa H, and
  Reznik B.
\newblock {Trapping of topological-structural defects in Coulomb crystals}.
\newblock {\em Phys. Rev. Lett.}, 110:133004, 2013.

\bibitem{Pyka2013a}
Pyka K, Keller J, Partner~H L, Nigmatullin R, Burgermeister T, Meier~D M,
  Kuhlmann K, Retzker A, Plenio~M B, Zurek~W H, del Campo~A, and
  Mehlst{\"a}ubler~T E.
\newblock {Symmetry breaking and topological defect formation in ion Coulomb
  crystals}.
\newblock {\em arXiv:1211.7005}, 2012.

\bibitem{Ulm2013}
Ulm S, Rossnagel J, Jacob G, Deg{\"u}nther C, Dawkins~S T, Poschinger~U G,
  Nigmatullin R, Retzker A, Plenio~M B, Schmidt-Kaler F, and Singer K.
\newblock {Observation of the Kibble-Zurek scaling law for defect formation in
  ion crystals}.
\newblock {\em arXiv:1302.5343}, 2013.

\bibitem{Ejtemaee2013}
Ejtemaee S and Haljan~P C.
\newblock {Spontaneous nucleation and dynamics of kink defects in zigzag arrays
  of trapped ions}.
\newblock {\em arXiv:1303.6723}, 2013.

\bibitem{Landa2010}
Landa H, Marcovitch S, Retzker A, Plenio~M B, and Reznik B.
\newblock Quantum coherence of discrete kink solitons in ion traps.
\newblock {\em Phys. Rev. Lett.}, 104:043004, Jan 2010.

\bibitem{Fishman2008}
Fishman S, De~Chiara G, Calarco T, and Morigi G.
\newblock Structural phase transitions in low-dimensional ion crystals.
\newblock {\em Phys. Rev. B}, 77:064111, Feb 2008.

\bibitem{Retzker2008}
Retzker A, Thompson~R C, Segal~D M, and Plenio~M B.
\newblock Double well potentials and quantum phase transitions in ion traps.
\newblock {\em Phys. Rev. Lett.}, 101:260504, Dec 2008.

\bibitem{Schneider2012}
Schneider C, Porras D, and Schaetz T.
\newblock Experimental quantum simulations of many-body physics with trapped
  ions.
\newblock {\em Reports on Progress in Physics}, 75(2):024401, 2012.

\bibitem{Nigmatullin2011}
Nigmatullin R, del Campo~A, de~Chiara~G, Morigi G, Plenio~M B, and Retzker A.
\newblock {Formation of helical ion chains}.
\newblock {\em arXiv:1112.1305}, 2011.

\bibitem{LandaPC}
Landa H.
\newblock private communication.

\bibitem{Kibble1980}
Kibble T~W B.
\newblock Some implications of a cosmological phase transition.
\newblock {\em Physics Reports}, 67(1):183 -- 199, 1980.

\bibitem{Zurek1996}
Zurek~W H.
\newblock Cosmological experiments in condensed matter systems.
\newblock {\em Physics Reports}, 276(4):177 -- 221, 1996.

\bibitem{delCampo2013}
del Campo~A, Kibble T~W B, and Zurek~W H.
\newblock {Causality and non-equilibrium second-order phase transitions in
  inhomogeneous systems}.
\newblock {\em arXiv:1302.3648}, 2013.

\bibitem{Braun2004}
Braun~O M and Kivshar~Y S.
\newblock {\em The Frenkel-Kontorova model, concepts, methods, and
  applications}.
\newblock Springer, 2004.

\bibitem{Pyka2013b}
Pyka K, Herschbach N, Keller J, and Mehlst{\"a}ubler~T E.
\newblock {A high-precision rf trap with minimized micromotion for an In$^+$
  multiple-ion clock}.
\newblock {\em arXiv:1206.5111}, 2012.

\bibitem{Steane1997}
Steane~A M.
\newblock The ion trap quantum information processor.
\newblock {\em Applied Physics B: Lasers and Optics}, 64(6):623--643, 1997.

\bibitem{Rosenau2003}
Rosenau P.
\newblock Hamiltonian dynamics of dense chains and lattices: or how to correct
  the continuum.
\newblock {\em Physics Letters A}, 311(1):39--52, 2003.

\bibitem{Metcalf1999}
Metcalf~H J and Van der Straten~P.
\newblock {\em Laser cooling and trapping}.
\newblock Springer Verlag, 1999.

\bibitem{Ghosh1995}
Ghosh~P K.
\newblock {\em Ion Traps}.
\newblock Oxford University Press, 1995.

\bibitem{Sugiyama1995}
Sugiyama K and Yoda J.
\newblock {Disappearance of Yb$^+$ in excited states from rf trap by background
  gases}.
\newblock {\em Jpn. J. Appl. Phys.}, 34:L584--L586, 1995.

\bibitem{Sugiyama1997}
Sugiyama K and Yoda J.
\newblock {Production of YbH$^{+}$ by chemical reaction of Yb$^{+}$ in excited
  states with H$_ {2}$ gas.}
\newblock {\em Phys. Rev. A}, 55:10--13, 1997.

\bibitem{Rutkowski2011}
Rutkowski~P X, Michelini~M C, Bray~T H, Russo N, Marçalo J, and Gibson~J K.
\newblock Hydration of gas-phase ytterbium ion complexes studied by experiment
  and theory.
\newblock {\em Theoretical Chemistry Accounts}, 129(3-5):575--592, 2011.

\bibitem{delCampo2010}
{del Campo A and De Chiara G and Morigi G and Plenio M B and Retzker A}.
\newblock {Structural defects in ion chains by quenching the external
  potential: The inhomogeneous Kibble-Zurek mechanism}.
\newblock {\em Phys. Rev. Lett.}, 105:075701, Aug 2010.

\bibitem{deChiara2010}
De~Chiara G, del Campo~A, Morigi G, Plenio~M B, and Retzker A.
\newblock Spontaneous nucleation of structural defects in inhomogeneous ion
  chains.
\newblock {\em New Journal of Physics}, 12(11):115003, 2010.

\bibitem{Zurek2009}
Zurek~W H.
\newblock {Causality in condensates: gray solitons as relics of BEC formation}.
\newblock {\em Physical review letters}, 102(10):105702, 2009.

\bibitem{Sabbatini2012}
Sabbatini J, Zurek~W H, and Davis~M J.
\newblock {Causality and defect formation in the dynamics of an engineered
  quantum phase transition in a coupled binary Bose Einstein condensate}.
\newblock {\em New Journal of Physics}, 14(9):095030, 2012.

\bibitem{PhysRevD.11.2943}
Gervais J-L and Sakita B.
\newblock {Extended particles in quantum field theories}.
\newblock {\em Phys. Rev. D}, 11(10):2943--2945, May 1975.

\bibitem{PhysRevD.12.1678}
Tomboulis E.
\newblock {Canonical quantization of nonlinear waves}.
\newblock {\em Phys. Rev. D}, 12(6):1678--1683, Sep 1975.

\bibitem{PhysRevB.38.6713}
Boesch R, Stancioff P, and Willis~C R.
\newblock {Hamiltonian equations for multiple-collective-variable theories of
  nonlinear Klein-Gordon equations: A projection-operator approach}.
\newblock {\em Phys. Rev. B}, 38(10):6713--6735, Oct 1988.

\bibitem{PhysRevB.33.1904}
Willis C, El-Batanouny M, and Stancioff P.
\newblock {Sine-Gordon kinks on a discrete lattice. I. Hamiltonian formalism}.
\newblock {\em Phys. Rev. B}, 33(3):1904--1911, Feb 1986.

\bibitem{PhysRevB.33.1912}
Stancioff P, Willis C, El-Batanouny M, and Burdick S.
\newblock {Sine-Gordon kinks on a discrete lattice. II. Static properties}.
\newblock {\em Phys. Rev. B}, 33(3):1912--1920, Feb 1986.

\bibitem{PhysRevB.39.361}
Boesch R and Willis~C R.
\newblock {Exact determination of the Peierls-Nabarro frequency}.
\newblock {\em Phys. Rev. B}, 39(1):361--368, Jan 1989.

\bibitem{PhysRevB.40.2284}
Boesch R, Willis~C R, and El-Batanouny M.
\newblock {Spontaneous emission of radiation from a discrete Sine-Gordon kink}.
\newblock {\em Phys. Rev. B}, 40(4):2284--2296, Aug 1989.

\end{thebibliography}

\end{document}